\documentclass[12pt,preprint]{emulateapj}
\shorttitle{Iron in DA White Dwarfs}
\shortauthors{Vennes et al.}

\journalinfo{To appear in the Dec 10, 2006 issue of ApJ}
\begin{document}

\title{Iron in Hot DA White Dwarfs}

\author{St\'ephane Vennes\altaffilmark{1}, Pierre Chayer$^{2,}$\altaffilmark{3}, Jean Dupuis\altaffilmark{4}, and Thierry Lanz\altaffilmark{5}}

\altaffiltext{1}{Department of Physics and Space Sciences, Florida Institute of Technology, 150 W University Blvd., Melbourne, FL 32901-6975}

\altaffiltext{2}{Department of Physics and Astronomy, Johns Hopkins University,
3400 N. Charles St., Baltimore, MD 21218-2686}

\altaffiltext{3}{Also at Department of Physics and Astronomy, University of Victoria, P.O. Box 3055, Station Csc, Victoria, BC V8W 3P6, Canada.}

\altaffiltext{4}{Canadian Space Agency, 6767 route de l'Aeroport, Saint-Hubert, QC J3Y 8Y9, Canada.}

\altaffiltext{5}{Department of Astronomy, University of Maryland, College Park, MD 20742.}

\email{svennes@fit.edu, chayer@pha.jhu.edu,  Jean.Dupuis@space.gc.ca, tlanz@umd.edu.}

\begin{abstract}
We present a study of the iron abundance pattern in hot hydrogen-rich (DA) white dwarfs. The study is based on new and archival far ultraviolet  spectroscopy of a sample of white dwarfs in the temperature range $30,000 {\rm K} \la T_{\rm eff} \la 64,000$K.  The spectra obtained with the {\it Far Ultraviolet Spectroscopic Explorer} along with spectra obtained with the {\it Hubble Space Telescope}  Imaging Spectrograph and the {\it International Ultraviolet Explorer}  sample \ion{Fe}{3} to \ion{Fe}{6} absorption lines enabling a detailed iron abundance analysis over a wider range of effective temperatures than previously afforded. The measurements reveal abundance variations in excess of two orders of magnitude between the highest and the lowest temperatures probed, but also show considerable variations (over one order of magnitude) between objects with similar temperatures and surface gravities. Such variations in cooler objects may be imputed to accretion from unseen companions or so-called circumstellar debris although the effect of residual mass-loss and selective radiation pressure in the hottest objects in the sample remain dominant.
\end{abstract}

\keywords{stars: abundances --- stars: atmospheres-- white dwarfs}

\section{Introduction}

The cooling history of white dwarfs involves phases of chemical transformation of their atmospheres. One such transformation occurs early on the hydrogen-rich (DA) cooling sequence. Far ultraviolet (FUV) spectroscopy of hot, hence young DA white dwarfs show that iron is a dominant source of atmospheric opacity \citep{ven1992, hol1993}. The effect of iron-group opacities has also been observed in extreme ultraviolet (EUV) spectra of the same stars as well as in other objects with similar parameters \citep{dup1995,wol1998,sch2002}. Before reaching a cooling age of a few million years ($t_{cool} \leq 2-3\times 10^6$ years) this source of opacity vanishes in most DA white dwarfs, and the metallicity (i.e., $n(Fe)/n(H)$ by number) drops from nearly solar values \citep{ven2001} to less than $10^{-2} (Fe/H)_{\odot}$, i.e., below the detection threshold of high-dispersion {\it International Ultraviolet Explorer} ({\it IUE}) or even Space Telescope Imaging Spectrograph (STIS) high-resolution spectra \citep{bar2003}. However, the absence of an effective iron abundance diagnostics in cooler white dwarfs ($20,000{\rm K} \la T_{\rm eff} \la 40,000$K, or $t_{cool} = 80-4 \times10^6$ years), reserved detailed iron abundance analyses to a few young objects ($t_{cool} \leq 2-3\times 10^6$ years). A spectral window limited to $\lambda > 1150$\AA\ restricts the iron abundance analysis to  \ion{Fe}{4}, \ion{Fe}{5}, and \ion{Fe}{6} species which are dominant in white dwarfs with  $T_{eff} \ga 40,000$K, and to \ion{Fe}{2} represented by a strong doublet at $\lambda\lambda$ 1260.533/1267.422\AA\ and dominant in cool white dwarfs ($T_{eff} \la 20,000$K). However, in the temperature range $20,000-40,000$ K the ionization balance shifts toward \ion{Fe}{3}. Spectral lines of \ion{Fe}{3} are prominent near $\lambda \approx 1125$ \AA\  in a spectral window made accessible by the {\it Far Ultraviolet Spectroscopic Explorer} ({\it FUSE}).

The new possibilities offered by {\it FUSE} were demonstrated in a study of the DA white dwarf GD 394 ($T_{eff}\approx37,000$ K). The presence of trace elements in the otherwise hydrogen-dominated atmosphere of GD 394 had been established, among others, by \citet{pae1989} in their analysis of EUV photometry of several hot white dwarfs observed with {\it EXOSAT}. The source of opacity remained uncertain \citep[see][]{wol1998, dup2000} until \citet{cha2000} measured a surprisingly high abundance of iron in their analysis of new {\it FUSE} spectra.  They used several lines of \ion{Fe}{3} to determine an iron abundance almost solar ($Fe/H \sim 5\times 10^{-6}$). Such a high abundance goes against the trend inferred in the study of hotter white dwarfs and indicate a resurgence of heavy elements in the atmosphere of aging DA white dwarfs. Since \citet{cha1995} demonstrated that the abundance of iron supported by selective radiation pressure decreases well under a solar abundance at $T_{eff} \la 50,000$ K,  external causes are suspected, in particular accretion from the ISM, a close companion, or from a circumstellar shell.

To investigate the question further, we initiated a study of FUV spectra of a sample of twelve DA white dwarfs  observed with {\it FUSE},  {\it IUE}, and STIS. The sample covers a range of effective temperatures from $\le 64,000$K down to $\ge 30,000$K. We describe the sample selection process  in \S 2, with relevant EUV/soft X-ray data listed in Appendix~A, and we present  the FUV observations in \S 3. The model atmospheres and spectral synthesis used in this investigation are presented in \S 4, and our analysis is presented in \S 5. Finally, we summarize and discuss our results in \S 6. 

\section{Sample selection: serendipitous and deliberate}

We searched for high-metallicity white dwarfs in the sample collected as part of the {\it FUSE} Z903   and P204 programs \citep{dup2005, ven2005}. The programs assembled over 40 DA white dwarfs selected for observations on the basis of their predicted UV luminosity and high equatorial  latitude. The sample presents  opportunities for the serendipitous identification of high metallicity DA white dwarfs. Next, we included three stars with large iron abundances initially studied by \citet{hol1993} and more recently  by \citet{bar2003}.

In a systematic effort to identify high-metallicity white dwarfs, we have also examined the EUV/soft X-ray properties of a large sample of 178  DA white dwarfs with effective temperatures ranging from 84,000 K down to 24,000 K, i.e., with corresponding ages ranging from 0.5 to 30 million years. The sample comprises isolated white dwarfs \citep{ven1996,ven1997,ven1999,ven2006} and in binaries with luminous companions \citep{ven1998}. Among these, some 104 objects are joint  detections from the {\it EUVE}/100\AA\ survey and the {\it ROSAT} PSPC survey \citep{bow1996,vog1999}.

\begin{figure}[!hb]
\plotone{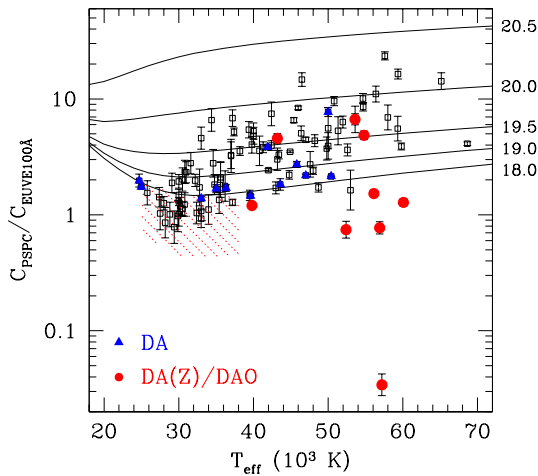}
\caption{Ratio of observed $ROSAT$/PSPC to $EUVE$/100\AA\ count rates as a function of effective temperatures for a sample of 104 objects. The data are compared to pure-H model predictions with varying ISM column densities ({\it full lines}) at  $\log n_H = 18, 18.5, 19, 19.5, 20, 20.5$. Note the spectroscopically confirmed  pure-H ({\it blue triangles}) and high-metallicity stars ({\it red circles}).}\label{fig1}
\end{figure}

Figure~\ref{fig1} shows the {\it ROSAT}/PSPC to {\it EUVE}/100\AA\ count rate ratio as a function of effective temperatures for spectroscopically-confirmed pure-H white dwarfs ({\it blue triangles}), and spectroscopically-confirmed high-metallicity DA/DAO white dwarfs ({\it red circles}). The remaining stars are shown with {\it open black squares}. The PSPC soft X-ray measurements probe the high-energy tail of a hot white dwarf soft X-ray emission ($\lambda \la 100$\AA), and the {\it EUVE}/100\AA\  measurements probe somewhat longer wavelength emission. The data, available in Appendix~A are compared to predictions from pure-hydrogen models at $\log{g}=8$ and varying intervening ISM column density expressed in number of neutral hydrogen atoms per cm$^2$ ({\it full lines} at $\log n_H = 18, 19, 19.5, 20, 20.5$). The minimum flux ratio expected from a pure-H atmosphere quickly converges  to values independent of the ISM column density as it approaches log $n_H \approx 18$. Values observed below the curve labelled log $n_H = 18$ are incompatible with pure hydrogen atmospheres and correspond to stars contaminated with heavy elements, possibly iron. Note that the effect of increased metallicity opposes the effect of increased ISM column density, therefore the method will select in preference objects with high metallicity and low neutral hydrogen column density in the ISM. The behavior of the count rate ratio for the sub-sample of pure-H white dwarfs ---among these are Sirius~B, GD~71, and HZ~43---and in particular near the temperature of 30,000K, guarantees that the procedure is well calibrated. Note that surface gravity variations are imperceptible. 

The {\it red-hatched} area encloses a particularly interesting sample of objects, and, immediately to the right of this area, the {\it red circle}  represents the high-metallicity DA GD~394. The grouping of stars found within the {\it red-hatched} area close to the high-metallicity white dwarf GD~394 suggests possible  spectroscopic similarities between these  objects. 

Therefore, to demonstrate that iron opacities are responsible, at least in part, for the anomalous flux ratios, we searched bright candidates with count rate ratios significantly below the minimum threshold. The DA white dwarfs GD~683 and LB~1663 \citep{ven1996} stood out with  PSPC to {\it EUVE} 100\AA\ flux ratios of $\sim 0.98$ and $\sim 0.93$, respectively. These ratios are inferior to the ratio measured  in GD~394 ($\sim 1.20$) and indicate the likely presence of heavy elements. These two objects have been observed as part of the {\it FUSE} D023 observing program. 

\section{FUV Spectroscopy}

\begin{figure}[ht!]
\plotone{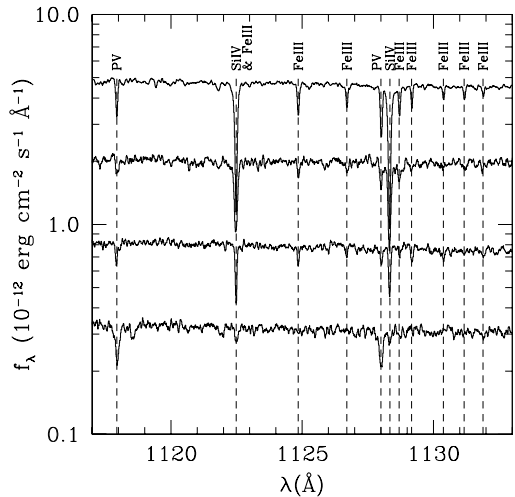}
\caption{$FUSE$ spectra showing iron line identifications in, from top to bottom, GD~394, WD~1819$+$580, GD~683, and WD~2000$-$561 (offset by $-0.3$dex). The iron line identifications in WD~J2000$-$561 are uncertain. The spectra also show  silicon and phosphorus lines.}\label{fig2}
\end{figure}

Table~\ref{tbl1} lists the sample of white dwarfs observed spectroscopically with {\it FUSE}. The observations were made using the $30\times30\arcsec$ low-resolution slit (LWRS) resulting in 100\% throughput and insuring complete spectral coverage ($900\la \lambda \la 1180$\AA) with a spectroscopic resolving power $R=\lambda/\Delta\lambda=20000\pm2000$. The stars 0549$+$158 (GD~71), 1819$+$580 (EUVE~J1820$+$58.0, RE~J1820$+$580) and 2000$-$561 (EUVE~J2004$-$56.0, RE~J2004$-$560) were selected from the Z903 and P204 programs, and the stars GD~683 and LB~1663 were observed as part of the D023 program. We also re-analyzed the {\it FUSE} spectra of the DA white dwarf GD~394 \citep{cha2000}. All data were processed with the {\it FUSE} data reduction pipeline CALFUSE version 3.0. Figure~\ref{fig2} shows \ion{Fe}{3} line identifications in {\it FUSE} spectra along with \ion{P}{5} and \ion{Si}{4} line identifications.

\begin{deluxetable*}{lclllcccc}
\tabletypesize{\scriptsize}
\tablecaption{{\it FUSE} Observations\label{tbl1}}
\tablewidth{0pt}
\tablehead{\colhead{WD name} & \colhead{other name} & \colhead{1RXS J} & \colhead{$T_{eff}$ (K)} & \colhead{log $g$} & \colhead{V (mag)\tablenotemark{a} }  & \colhead{Data ID} & \colhead{$t_{\rm exp}$ (s)} & \colhead{Aperture}\\
} 
\startdata
0106$-$358 & GD~683, EUVE, RE & 010821.4$-$353433 & $29000\pm600$\tablenotemark{b} & $7.99\pm0.25$\tablenotemark{b} & 14.70  & D0230101 & 30835 & LWRS \\ 
0320$-$539 & LB~1663, EUVE, RE & 032215.5$-$534515  & $32880\pm1070$\tablenotemark{b} & $7.98\pm0.16$\tablenotemark{b}    &   14.90 & D0230201 & 9062   & LWRS \\
0549$+$158 & GD~71, EUVE, RE   & 055228.1$+$155313 &  $33000\pm300$\tablenotemark{c} & $7.87\pm0.15$\tablenotemark{c}   & 13.03  & P2041701 &   13929   & LWRS \\
1819$+$580 & EUVE, RE & 182030.0$+$580437 & $44200\pm700$\tablenotemark{c} & $8.61\pm0.21$\tablenotemark{c} & 13.95 &  Z9032801 & 4698 & LWRS  \\
2000$-$561 & EUVE, RE  & 200425.2$-$560301 & $42200\pm200$\tablenotemark{c} & $7.84\pm0.18$\tablenotemark{c} & 14.97  &  Z9033301 &  11596 & LWRS \\
2111$+$498 & GD 394, EUVE, RE   & 211244.1$+$500616 & $36700\pm400$\tablenotemark{c} & $8.27\pm0.12$\tablenotemark{c} & 13.08  &  P1043601 & 28635 & LWRS \\
\enddata
\tablenotetext{a}{From \citet{mcc1999} and references therein.}
\tablenotetext{b}{This work.}
\tablenotetext{c}{From \citet{ven2005}.}
\end{deluxetable*}
 
\subsection{STIS and {\it IUE} Observations}

\begin{deluxetable}{lcc}
\tablecaption{STIS and $IUE$ Observations\label{tbl2}}
\tablewidth{0pt}
\tablehead{\colhead{WD name} & \colhead{Data ID} & \colhead{$t_{\rm exp}$ (s)} \\
}
\startdata
0556$-$375 &  O59P03010   &  2494 \\
                        & O59P03020   &  2977 \\
                         & O59P03030   &  2977 \\
0621$-$376 &  SWP45951HL\tablenotemark{a} & 14400 \\ 
                       &   SWP49037HL  &   13800  \\
                        &   SWP49038HL  &  12600  \\
                        &   SWP49039HL  &  12600   \\
2211$-$495 &   SWP44766HL & 7200    \\
                        &     SWP44767HL   & 7200   \\
                        &     SWP47954HL    &  7200   \\
                         &    SWP47955HL     & 7200  \\
                         &    SWP47956HL     & 7200  \\
                         &     SWP47996HL     &  7200  \\
\enddata
\tablenotetext{a}{SWP=short wavelength primary; HL = high dispersion large aperture.}
\end{deluxetable}

We obtained a series of STIS spectra of the hot white dwarf WD0556$-$375 (EUVE~J0558$-$37.5, RE~J0558$-$376) from the Multi-Mission Archive at the Space Telescope (MAST, Table \ref{tbl2}). The spectra were obtained in the $0.2\times 0.2\arcsec$ aperture and using the E140M grating resulting in
a resolving power of $R\sim40000$. We also obtained a series of {\it IUE} high-dispersion spectra of the hot DA white dwarfs WD0621$-$376 (EUVE~J0623$-$37.6, RE~J0623$-$374) and WD2211$-$495 (EUVE~J2214$-$49.3, RE~J2214$-$492) from MAST (Table \ref{tbl2}). The spectra were obtained in the high-dispersion mode and the large aperture resulting in a resolving power $R\sim10000$ and a wavelength coverage of $1150\la \lambda \la 1950$\AA. We co-added the spectra for each star.

\section{Model Atmospheres and Spectra}

The model atmospheres and synthetic spectra were calculated assuming non-local thermodynamic (non-LTE) equilibrium using the fortran codes TLUSTY version 200  and SYNSPEC version 48 \citep{hub1995}. Using these codes, \citet{ven2005} computed a grid of pure hydrogen model atmospheres covering the range of effective temperatures from $T_{\rm eff} = 20,000$K to 80,000 K and surface gravities from $\log{g}=6.8$ to 9.6. By comparing a grid computed assuming local thermodynamic equilibrium (LTE) to the non-LTE grid, \citet{ven2005} demonstrated that non-LTE effects in the Lyman line spectra of DA white dwarfs are for the most part negligible.  On the other hand, \citet{ven2005} also studied the effect of heavy element opacities on the Lyman and Balmer line spectra of high-metallicity white dwarfs such as G191-B2B and 0621$-$376. They demonstrated that the observed abundance of heavy elements in the photospheres of these stars accounts for discrepancies noted in effective temperatures and surface gravities measured with Balmer line spectra versus the same parameters measured with Lyman line spectra. The models assume homogeneous distribution of heavy elements over the surface, and as a function of depth in the atmosphere. Note that \citet{sch2002} explicitly considered a vertical distribution of heavy elements in diffusive equilibrium. A basic feature of their ab initio calculations is that stars with similar effective temperatures and surface gravities ought to display similar abundance patterns. This is not always the case as we shall demonstrate in \S 5, and
equilibrium scenarios may easily be disrupted by accretion from a companion or even a weak wind.

Based on these considerations, we created three sets of models for the purpose of (1) analyzing the Lyman line spectra of cooler objects (GD~71, GD~683, LB~1663, 1819$+$580, 2000$-$561, and GD~394), (2) measuring the trace element abundances in the same objects, and, finally, (3) measuring the abundance of trace elements in the high-metallicity DA white dwarfs 0621$-$376, 2211$-$495, and 0556$-$375.

\begin{figure}[ht]
\plotone{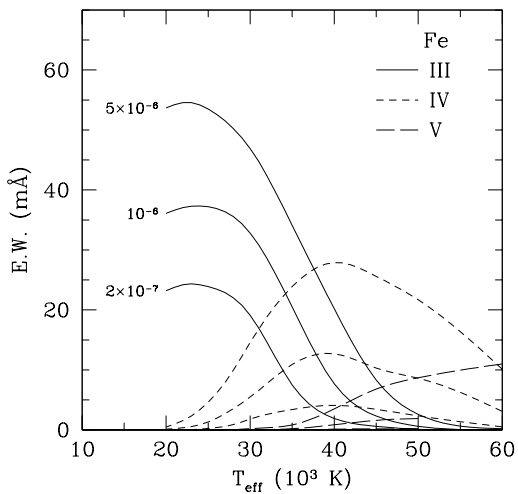}
\caption{Equivalent widths of \ion{Fe}{3}$\lambda$1124.881, \ion{Fe}{4}$\lambda$1542.698, and \ion{Fe}{5}$\lambda$1376.337 as a function of effective temperatures. \ion{Fe}{3}$\lambda$1124.881 is an excellent iron abundance diagnostics for temperatures below 30,000K, and the line remains detectable up to a temperature of $\approx$45,000K. }\label{fig3}
\end{figure}

To analyze the Lyman line spectra obtained with {\it FUSE} (Table \ref{tbl1}) we adopted the LTE model grid of \citet{ven2005}. The Lyman line profiles are computed using the tables of \citet{lem1997}. Next, having obtained the temperature and surface gravity of each star, we computed sets of three non-LTE models for each star with model atoms from the OSTAR2002 series \citep{lan2003}. The stellar temperatures within this particular group range from $\approx 30,000$ to $\approx 45,000$K, therefore we include the ions \ion{H}{1} (9 levels), \ion{H}{2}, \ion{C}{2} (22 levels), \ion{C}{3} (23 levels), \ion{C}{4} (25 levels), \ion{C}{5} (1 level), \ion{Si}{3} (30 levels), \ion{Si}{4} (23 levels), \ion{Si}{5} (1 level), \ion{P}{4} (14 levels), \ion{P}{5} (17 levels), \ion{P}{6} (1 level), \ion{Fe}{2} (36 super-levels), \ion{Fe}{3} (50 super-levels), \ion{Fe}{4} (43 super-levels), \ion{Fe}{5} (42 super-levels), and \ion{Fe}{6} (1 level). Super-levels group together levels with the same basic configuration. Figure \ref{fig3} shows the equivalent widths of selected iron lines as a function of effective temperatures for abundances of $\log{\rm Fe/H} = -5.3, -6.0$, and $-6.7$. The resonance lines of \ion{Fe}{3} are dominant at effective temperatures $T_{\rm eff} \la 40,000$K. 

Next, we extended the sequence of models presented by \citet{ven2001} in their study of Feige~24 and G191-B2B to higher abundances and temperatures encompassing likely parameters for the hot, high-metallicity DA white dwarfs 0556$-$375, 0621$-$376, and 2211$-$495. Three new models at $T_{\rm eff} = 57, 60, 64\times10^3$K and $\log{g}=7.5$ share a high heavy-element concentration of $\log{\rm He/H}=-5.0$, $\log{\rm C/H}=-6.0$, $\log{\rm N/H}=-5.8$, $\log{\rm O/H}=-5.0$, $\log{\rm Si/H}=-5.4$, $\log{\rm S/H}=-5.0$, $\log{\rm Fe/H}=-4.0$, and $\log{\rm Ni/H}=-4.8$, while three other models at $T_{\rm eff} = 57, 60, 64\times10^3$K ($\log{g}=7.5$) share a lower concentration used by \citet{ven2001}. Together, these six new models cover the range of abundances and effective temperatures appropriate for an abundance study of 0556$-$375, 0621$-$376 and 2211$-$495.

\section{Analysis and Results}

We assembled twelve stars with temperatures ranging from $\approx 30,000$ to $\approx 64,000$K and we determined the abundance of iron in their photospheres. Two stars were observed with {\it IUE} (0621$-$376 and 2211$-$495) while another was observed with STIS, and  six stars were observed with {\it FUSE} (Table \ref{tbl1}).  To this sample we added published iron abundance analyses of the hot DA white dwarfs Feige~24, G191-B2B, and GD~246. Incidentally, we also measured abundances of carbon, silicon and phosphorous in the sample of stars observed with {\it FUSE}; these specific results will be discussed at a later time.
 
\subsection{Feige~24, G191-B2B, and GD~246}

In the cases of Feige~24 and G191-B2B, we adopted the abundance analysis of \citet{ven2001}. Their analysis is based on the same generation of iron-rich models used in the present study.
The iron equivalent widths in Feige~24 and G191-B2B are very similar and the iron abundances  in these two stars are very close. For example, the  \ion{Fe}{4}$\lambda1542.698$ equivalent width is 11~m\AA\ in both objects, and the \ion{Fe}{5}$\lambda1370.337$ equivalent width in Feige~24 and G191-B2B is 26 and 22 m\AA, respectively. We adopted temperatures of $T_{\rm eff} = 57,000$K and 55,000K for Feige 24 and G191-B2B, respectively, as estimated by  \citet{ven2001}. The corresponding iron abundances are $\log{\rm Fe/H} = -5.5$ and $-5.6$ for Feige 24 and G191-B2B, respectively. In the case of GD~246 we adopted the analysis of a {\it Chandra} Low Energy Transmission Grating \citep{ven2002}. The abundance analysis of GD~246 was based on the detection of \ion{Fe}{5}, \ion{Fe}{6} and \ion{Fe}{7} lines and resulted in an iron abundance estimate of $\log{\rm Fe/H}\approx -6.5$ and $T_{\rm eff}=54,000$K.

\subsection{Analysis of STIS and {\it IUE} Spectroscopy}

\begin{figure}[hb]
\plotone{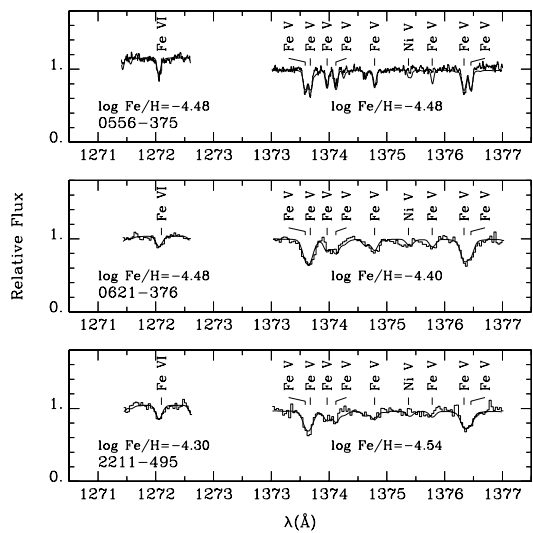}
\caption{{\it IUE} ultraviolet spectra of  WD0621$-$376 and WD2211$-$495 ($R\approx10,000$), and STIS spectrum ($R\approx40,000$) of WD0556$-$375, and best \ion{Fe}{5} and \ion{Fe}{6} line profile fits.}\label{fig4}
\end{figure}

Figure \ref{fig4} shows the iron abundance analysis of the FUV spectra of the hot DA white dwarfs 0556$-$375, 0621$-$376, and 2211$-$495. We applied a $\chi^2$ minimization technique to determine the abundances and 1$\sigma$ errors. The analysis is restricted to a few well defined spectral lines with accurate $\log{gf}$. 

We adopted effective temperatures deduced from the Balmer and Lyman line analysis of \citet{ven2005}. \citet{ven2005} measured the effect of metallicity using non-LTE model atmospheres and deduced effective temperatures of $T_{\rm eff}=60,000$K and 62,000K for 0621$-$376 and 2211$-$495, respectively. Adopting the model grid  of \citet{ven2005}, we repeated the Balmer line analysis of 0556$-$375 \citep{ven1997} and measured $T_{\rm eff}=68800\pm1000$K and $\log{g}=7.56\pm0.05$. The DA white dwarf 0556$-$375 is hotter than 2211$-$495 by $\approx 2800$K and is hotter than 0621$-$376 by $\approx 4000$K. Again, using non-LTE metallicity vectors calculated by \citet{ven2005}, we adopted a temperature of $T_{\rm eff}=64000$K for 0556$-$375. Hence, the iron abundances based on FUV spectra are $\log{\rm Fe/H}=-4.44\pm0.10$, $-4.42\pm0.10$, and  $-4.48\pm0.10$ for 0621$-$376, 2211$-$495, and 0556$-$375, respectively. The abundances based on \ion{Fe}{5} lines are consistent with abundances based on \ion{Fe}{6} lines which implies that models calculated at the adopted stellar parameters predict the ionization balance correctly.

\subsection{Analysis of {\it FUSE} Spectroscopy}

\begin{figure}
\plotone{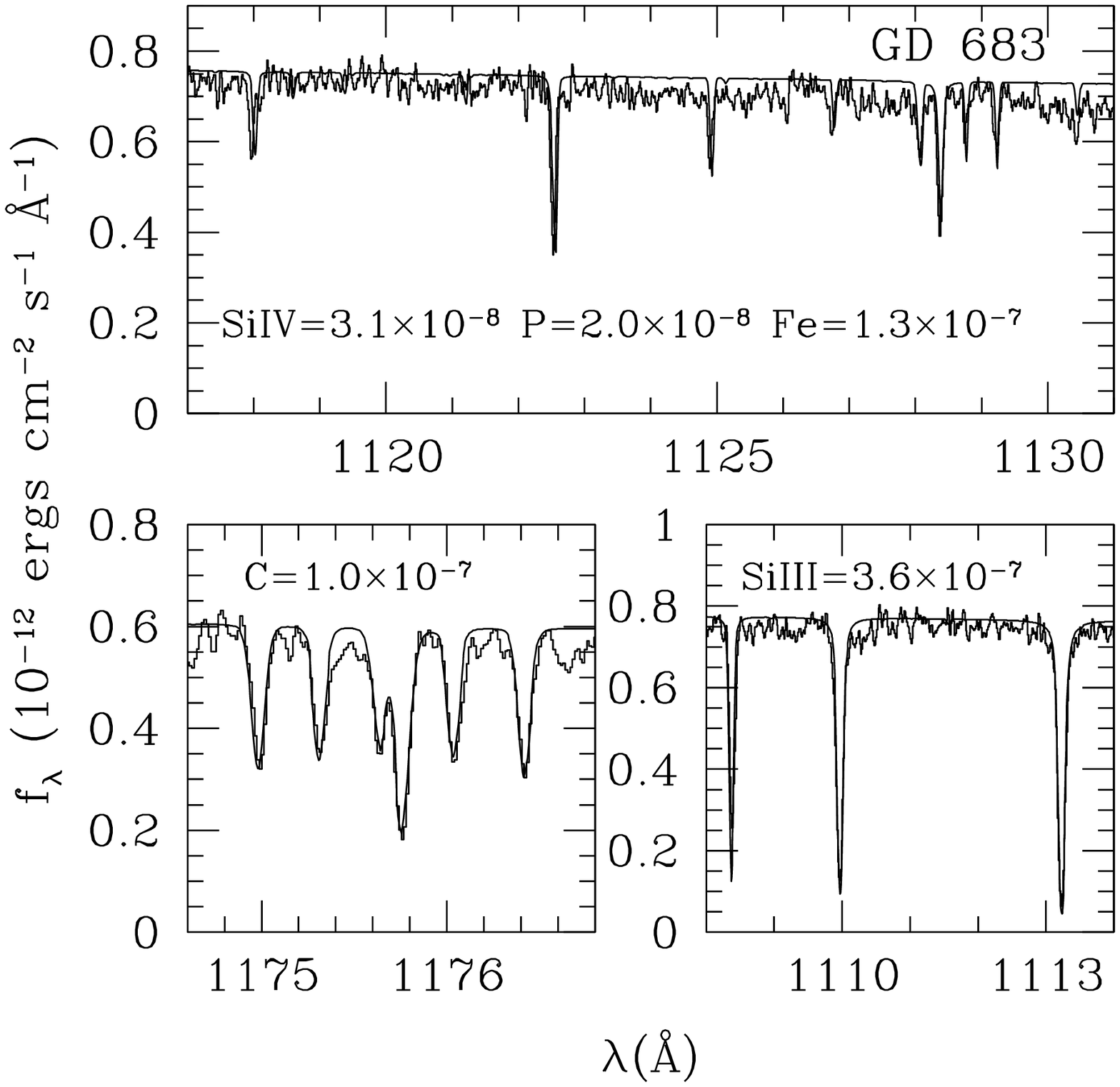}
\caption{Ultraviolet spectrum of GD 683 and best \ion{C}{3}, \ion{Si}{3}, \ion{Si}{4}, \ion{P}{5}, and \ion{Fe}{3} line profile fits.}\label{fig5}
\end{figure}

Again, the analysis is restricted to a few well defined spectral lines with accurate $\log{gf}$. The wavelength range covering the \ion{C}{3} line profile fits is $1174\le\lambda\le 1177$ with the continuum measured on each side. Similarly the wavelength ranges covering the \ion{Fe}{3}, \ion{P}{5}, \ion{Si}{3}, and \ion{Si}{4} are $1124\le\lambda\le 1127.5$, $1117\le\lambda\le 1119$, $1108 \le\lambda\le 1114$, and $1122 \le\lambda\le 1123$\AA, respectively. We applied a $\chi^2$ minimization technique to determine the abundances and 1$\sigma$ errors, or, when appropriate, the abundance upper limits. Table \ref{tbl3} summarizes our {\it FUSE}  abundance analysis. Figure \ref{fig5} shows the results of the analysis for the DA white dwarf GD~683. The intrinsic strengths of the \ion{Fe}{3} lines are sufficient to allow abundance measurements $\approx$1.5 dex lower than in the extreme case of GD~394. 

\begin{deluxetable*}{lccccc}
\tablecaption{Abundances Based on {\it FUSE} Spectroscopy\label{tbl3}}
\tablehead{\colhead{WD name} & \colhead{C/H} & \multicolumn{2}{c}{Si/H} & \colhead{P/H} & \colhead{Fe/H} \\
\cline{3-4}\\
                                                  &                            &    ( III)   &     (IV)   &                          &                              \\ 
                                                  }
\startdata
0106$-$358 &    $-7.00\pm0.07$     &  $-6.44\pm0.06$ & $-7.51\pm0.17$           &    $-7.69\pm0.17$        &   $-6.87\pm0.19$         \\
0320$-$539 &    $<-8.4$              &    $<-8.7$                     &  $<-8.3$                        &    $ <-8.7$                 &  $<-6.9$                       \\ 
0549$+$158 &   $<-8.9$             &    $<-9.0$                   &  $<-9.0$                           &  $-8.7\pm0.2$\tablenotemark{a}                  &   $<-7.3$                      \\
1819$+$580 &  $-7.24\pm0.07$      &  $-7.71\pm0.08$ &  $-8.44\pm0.11$          &    $-7.64\pm0.25$  &   $-5.91\pm0.29$              \\
2000$-$561 &  $-7.76\pm0.06$ & $-7.50\pm0.07$ & $-8.54\pm0.10$  &  $-7.84\pm0.14$ & $<-6.0$    \\
2111+498      &  $< -8.8$         &   $-5.67\pm0.04$ & $-6.38\pm0.07$   &  $-8.00\pm0.08$   &  $-5.36\pm0.09$      \\
\enddata
\tablenotetext{a}{\citet{dob2005} reported the detection of a week \ion{P}{5}$\lambda$1118\AA\  line in GD~71 and measured $\log{\rm P/H}=-8.6$ in agreement with our measurement.}
\end{deluxetable*}

The ionization balance of silicon is problematic  in all stars. The abundances based on \ion{Si}{3} lines are between 0.7 and 1.0 dex larger than abundances based on \ion{Si}{4} lines in the same stars. The causes of this ionization  imbalance, already noted by \citet{dup2000} in the case of GD~394, remain unknown. The present abundance measurements  of 1819$+$580 and 2000$-$561 are somewhat at variance with those reported by \citet{dup2005} which is in part due to the  more comprehensive atmospheric compositions used in the present study, and, in the the case of 1819$+$580 to the use of atmospheric parameters determined using the Lyman line series rather than the Balmer line series.

\subsection{The Iron Abundance Pattern}

\begin{figure}[h!b]
\plotone{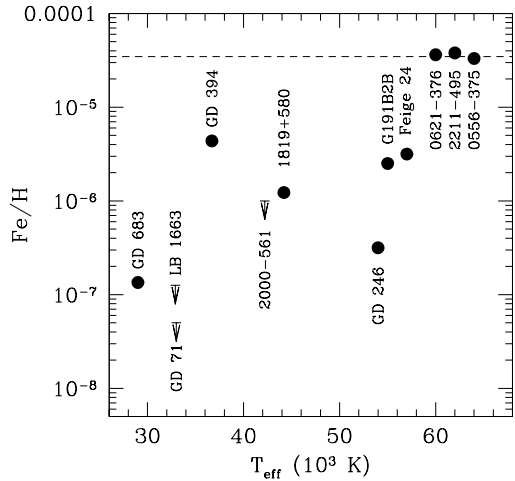}
\caption{Iron abundance in a sample of DA white dwarfs observed with {\it HST} (G191-B2B and Feige 24), {\it FUSE} (GD~71, GD~683, LB~1663, GD~394, 2000$-$561, and 1819$+$580), {\it IUE} (0621$-$376 and 2211$-$495), {\it Chandra} (GD 246), and STIS (0556$-$375). The horizontal dashed line shows the solar abundance.}\label{fig6}
\end{figure}

Figure \ref{fig6} shows the iron abundance measurements in our white dwarf selection as a function of effective temperatures. The white dwarfs GD~683 and LB~1663 are closely matched in effective temperatures, but a poorer signal-to-noise ratio in the latter only allowed to establish an upper limit to the iron abundance. Although the abundances of carbon, silicon and phosphorous are markedly lower in LB~1663, the presence of iron at a comparable level than in GD~683 cannot be excluded and its presumed effect on the EUV/soft X-ray emission of LB~1663 remains a likely explanation for the anomalous PSPC/{\it EUVE} 100\AA\ flux ratio which prompted its selection. On the other hand, stringent abundance upper limits were established in the case of bright DA white dwarf GD~71 showing that objects with similar effective temperatures and surface gravities, such as GD~71 and GD~683, may have dissimilar abundance patterns. The DA GD~71 is among those objects, depicted in Figure~\ref{fig1}, that are characterized by a very low metallicity.

The white dwarfs 1819$+$580 and 2000$-$560 are similarly paired and again a poorer signal-to-noise ratio in the latter does not permit to determine the iron abundance. However, the two stars show similar levels of carbon, silicon, and phosphorous. Clearly, 2000$-$561 should be re-observed with {\it FUSE} for a longer exposure time increasing the prospect of detecting iron in its photosphere. 

Our non-LTE analysis of GD~394 confirms the results obtained from the LTE analysis of \citet{cha2000}. In fact, the abundance of iron in the photosphere of GD~394 exceeds that of G191-B2B and Feige~24. The hottest, hence youngest DA white dwarfs in the sample also display the highest, almost precisely solar iron abundance. 

Figure \ref{fig6} demonstrates the chemical transformation occurring early on ($t_{\rm cool} \la 3\times10^6$ years) along the DA white dwarf cooling sequence. However, the obvious resurgence of iron in the photospheres of some  objects characterized by temperatures $T_{\rm eff} \la 45,000$ K is puzzling. Clearly, equilibrium diffusion calculations \citep{cha1995} do not explain this behavior, and contributions from iron reservoirs external to the stars may be required. 

\section{Discussion and Summary}

We have presented a study of the iron abundance pattern in hot DA white dwarfs. We reported the discovery of \ion{Fe}{3} spectral lines in a {\it FUSE} spectrum of the DA white dwarf GD~683 and a corresponding abundance $\log{\rm Fe/H}=-6.9$. The discovery of iron in the photosphere of this star confirms our suspicion that heavy elements, and iron in particular, are responsible for lingering EUV/soft X-ray flux deficit in DA white dwarfs with temperatures as low as $\approx$30,000K.   The detection of iron in the DA white dwarf 1819+580 is particularly intriguing. \citet{gre2000} uncovered substantial near infrared excess from this star which  they attributed to a dM6 companion, but there is insufficient information to establish its proximity to the white dwarf.

In fact, with the exception of Feige~24 and 1819+580 we do not find  clear evidence of an infrared flux excess in any of the sample stars. We obtained JHK photometry from the 2MASS database available at the Centre de Donnees astronomique de Strasburg and V magnitudes from \citet{mcc1999}, and we computed $V-J$ indices for all objects in the sample. The index ranges from $V-J=-0.72\pm0.10$ (GD~683) to $-0.96\pm0.15$ (2000$-$561) with the exception of 1819$+$580 ($V-J=-0.11\pm0.10$) and Feige~24 ($V-J=1.15\pm0.10$). Synthetic $V-J$ indices for DA white dwarfs range from $-0.7$ to $-0.8$ for white dwarf effective temperatures between 30,000K and 50,000K. Allowing for large uncertainties in 2MASS measurements, it appears that only 1819$+$580 and Feige~24 show evidence of a companion in infrared measurements. The estimated absolute J magnitudes are $M_J=11.43$ and 7.06  corresponding to M8.5 and M2.5 spectral types \citep{kir1994} for 1819$+$580 and Feige~24, respectively. The spectral type of the companion of 1819$+$580 is somewhat later than estimated by \citet{gre2000} because we adopted a higher gravity, hence lower luminosity, for the white dwarf itself. From our sample, only Feige~24 has a known {\it close} companion ($P=4.2$ days), but its abundance pattern is indistinguishable  from its sibling and apparently single white dwarf G191-B2B \citep{ven2001}. In all other cases, predicted $3\sigma$ upper limits to the companion absolute J magnitude range from $M_J \la10.3$ or a spectral type earlier than M6 \citep{kir1994} to $M_J\la 13.3$ or a  spectral type earlier than L5 \citep{kir2005}. These hypothetical stellar and  substellar  companions may be responsible for the presence of iron in white dwarfs such as GD~683 and GD~394. Moreover, the variable EUV emission from GD~394 may imply that iron, silicon and other elements are not uniformly distributed over the surface, which may be the result of episodic accretion from a putative companion \citep{dup2000}.

The iron abundance in DA white dwarfs with temperatures between $\approx30,000$ and $\approx 64,000$K does not correlate with effective temperatures in a simple manner.  The sample of DA white dwarfs is part of a EUV/soft X-ray selected population and may be biased against ultra-high metallicity objects, although the largest iron abundance measured is essentially solar. Other non-EUV selected white dwarfs were included in a study by \citet{bar2003} such as PG~0948$+$534, Ton~21, and PG~1342$+$444 which, although hotter than stars from the present study, may have lower iron abundances. Other EUV-selected DA white dwarfs such as 1RXS~J0619.1$-$0828 \citep{ven1999} lack FUV data required for an abundance analysis. Hopefully future missions may help enlarge this sample of hot DA white dwarfs.
 
In summary, our measurements reveal abundance variations in excess of two orders of magnitude between the highest and the lowest temperatures probed, but also show considerable variations (over one order of magnitude) between objects with similar temperatures and surface gravities. Such variations in cooler objects may be imputed to accretion from unseen companions or so-called circumstellar debris \citep{kil2005} although the effect of residual mass-loss and selective radiation pressure in the hottest objects in the sample may remain dominant. Accretion from a close companion remain a likely source of heavy elements in white dwarf photospheres. A deeper search for infrared excess, revealing the presence of a companion or debris near the white dwarf, may offer clues as to the source of iron in hot DA photospheres.

Guided by present and previous results we will embark on a re-analysis of the complete spectral energy distribution of all hot DA white dwarfs lifting the often convenient but possibly inappropriate assumption of a pure hydrogen composition. In particular, the present analysis of \ion{Fe}{3} spectral lines, and its sensitive iron abundance diagnostics, will be extended to the complete {\it FUSE} spectroscopic data set. 

\begin{acknowledgements}
We acknowledge support from a NASA LTSA grant (NAG5-11844) and a FUSE GI grant (NNG04GF05G). This research also received financial support from the College of Science of the Florida Institute of Technology. P.C. is a Canadian representative to the {\it FUSE} Project supported by CSA under a PWGSC contract. We thank A. Kawka for supplying infrared colors, and the referee
J. Holberg for useful comments.

This research has made use of the Simbad database and the VizieR service, operated at CDS, Strasbourg, France. This publication makes use of data products from the Two Micron All Sky Survey, which is a joint project of the University of Massachusetts and the Infrared Processing and Analysis Center, funded by the National Aeronautics and Space Administration and the National Science Foundation.
\end{acknowledgements}

\appendix
\section{Soft X-ray to EUV count rate ratios \label{appendix}}
Table~\ref{tbl4} lists the PSPC to {\it EUVE} 100\AA\ count rate ratios for a sample of 104 hot DA white dwarfs detected in both surveys \citep{bow1996, vog1999}. The stars are listed with the PSPC catalog designation, and  temperature estimates are provided in references cited in \S 2, except for 1RXS~J082704.9$+$284411 which is taken from \citet{heb1993}. These temperature estimates are used in the selection process depicted in Figure~\ref{fig1} only. Many have been updated since.

\begin{deluxetable*}{cccccc}
\tabletypesize{\scriptsize}
\tablecaption{Soft X-ray to EUV count rate ratios\label{tbl4}}
\tablewidth{0pt}
\tablehead{\colhead{{\it ROSAT} name} & \colhead{$C_{\rm PSPC}/C_{\rm 100\AA}$} & \colhead{$T_{\rm eff}$}  & \colhead{{\it ROSAT} name} & \colhead{$C_{\rm PSPC}/C_{\rm 100\AA}$} & \colhead{$T_{\rm eff}$}\\
                                                                        &                                              &   (10$^3$K)   &                                      &                                              &   (10$^3$K) \\
}
\startdata
1RXS J000359.1$+$433600 &   7.44$\pm$ 1.72 &  42.4 & 1RXS J103347.4$-$114146 &   1.55$\pm$ 0.37 &  25.8 \\
1RXS J000732.4$+$331730 &   9.58$\pm$ 0.87 &  50.8 & 1RXS J103624.9$+$460838 &   1.19$\pm$ 0.21 &  30.2 \\
1RXS J002958.0$-$632500 &   3.89$\pm$ 0.27 &  59.8 & 1RXS J104311.5$+$490227 &   2.40$\pm$ 0.17 &  48.0  \\
1RXS J005317.6$-$325951 &   1.71$\pm$ 0.11 &  36.3  & 1RXS J104446.6$+$574449 &   2.35$\pm$ 0.53 &  30.8 \\
1RXS J010821.4$-$353433 &   0.98$\pm$ 0.29 &  29.9 & 1RXS J105820.6$-$384416 &   0.85$\pm$ 0.25 &  28.2  \\
1RXS J013424.0$-$160708 &   1.72$\pm$ 0.16 &  48.7 & 1RXS J105916.6$+$512452 &   4.11$\pm$ 0.15 &  68.6  \\
1RXS J013853.1$+$252325 &   5.43$\pm$ 0.88 &  39.4 & 1RXS J110036.4$+$713758 &   1.71$\pm$ 0.20 &  43.0 \\
1RXS J015109.6$+$673947 &   2.33$\pm$ 0.48 &  31.0 & 1RXS J111140.4$-$224933 &   1.48$\pm$ 0.17 &  30.0\\
1RXS J022818.9$-$611817 &   4.44$\pm$ 0.22 &  47.0  & 1RXS J111238.8$+$240909 &   4.03$\pm$ 0.66 &  39.8\\
1RXS J023725.5$-$122129 &   1.04$\pm$ 0.22 &  32.4 & 1RXS J112619.1$+$183925 &   0.77$\pm$ 0.09 &  56.9 \\
1RXS J023947.9$+$500349 &   2.79$\pm$ 0.75 &  34.6 & 1RXS J114804.5$+$183042 &   1.03$\pm$ 0.32 &  27.5 \\
1RXS J030437.1$+$025706 &   2.02$\pm$ 0.68 &  35.6  & 1RXS J120055.5$-$363013 &   3.23$\pm$ 1.03 &  37.0 \\
1RXS J031713.9$-$853231 &   1.94$\pm$ 0.26 &  30.0  & 1RXS J123645.3$+$475530 &   1.53$\pm$ 0.08 &  56.1 \\
1RXS J032215.5$-$534515 &   0.93$\pm$ 0.16 &  33.0 & 1RXS J125702.4$+$220155 &   1.45$\pm$ 0.04 &  39.6 \\
1RXS J033714.6$-$415541 &   5.58$\pm$ 2.16 &  50.0  & 1RXS J131621.4$+$290555 &   2.15$\pm$ 0.01 &  50.4\\
1RXS J034850.1$-$005823 &   4.57$\pm$ 0.44 &  43.2  & 1RXS J144006.0$+$750539 &   3.96$\pm$ 0.37 &  42.4\\
1RXS J035629.1$-$364134 &   3.94$\pm$ 1.01 &  50.0  & 1RXS J144602.4$+$632929 &   3.56$\pm$ 1.02 &  40.8\\
1RXS J042734.0$+$740735 &  10.07$\pm$ 1.01 &  54.7 & 1RXS J150021.1$+$745838 &   5.55$\pm$ 1.33 &  59.3 \\
1RXS J044307.1$-$034655 &  14.22$\pm$ 2.34 &  65.1 & 1RXS J152944.2$+$483623 &   2.73$\pm$ 0.67 &  47.6  \\
1RXS J045712.5$-$280754 &   0.03$\pm$ 0.01 &  57.2  & 1RXS J153545.2$-$772439 &   6.93$\pm$ 1.67 &  58.0\\
1RXS J051206.1$-$004149 &   1.98$\pm$ 0.24 &  32.2  & 1RXS J162334.0$-$391359 &   1.97$\pm$ 0.25 &  24.7 \\
1RXS J051223.5$-$414525 &   3.63$\pm$ 0.43 &  52.6  & 1RXS J162909.4$+$780439 &   3.50$\pm$ 0.06 &  44.9\\
1RXS J051523.8$+$324107 &   2.43$\pm$ 0.08 &  42.0  & 1RXS J163825.8$+$350006 &   6.83$\pm$ 1.11 &  37.2\\
1RXS J052119.2$-$102925 &   1.08$\pm$ 0.21 &  33.0  & 1RXS J165851.0$+$341852 &   5.31$\pm$ 1.44 &  51.3 \\
1RXS J055038.1$+$000553 &   5.05$\pm$ 0.74 &  46.4 & 1RXS J165948.2$+$440059 &   1.13$\pm$ 0.15 &  30.5 \\
1RXS J055047.4$-$240853 &   6.66$\pm$ 1.72 &  53.6  & 1RXS J171127.2$+$664532 &  14.69$\pm$ 1.95 &  46.5\\
1RXS J055228.1$+$155313 &   1.39$\pm$ 0.06 &  33.0 & 1RXS J172642.8$+$583726 &  23.53$\pm$ 1.85 &  57.6  \\
1RXS J060502.5$-$481944 &   2.11$\pm$ 0.61 &  35.6  & 1RXS J172738.9$-$360015 &   1.72$\pm$ 0.60 &  32.8 \\
1RXS J063258.1$-$050547 &   3.30$\pm$ 0.28 &  43.4  & 1RXS J174614.9$-$703902 &   3.83$\pm$ 0.41 &  41.3\\
1RXS J063350.3$+$104122 &   1.42$\pm$ 0.27 &  27.4 & 1RXS J180009.9$+$683557 &   8.37$\pm$ 0.27 &  46.0 \\
1RXS J064509.3$-$164241 &   1.74$\pm$ 0.04 &  25.0 & 1RXS J182030.0$+$580437 &   6.55$\pm$ 0.33 &  45.4 \\
1RXS J064855.3$-$252350 &   1.26$\pm$ 0.16 &  28.0 & 1RXS J184509.8$+$682239 &   5.19$\pm$ 0.41 &  37.4 \\
1RXS J065414.2$-$020940 &   1.11$\pm$ 0.32 &  34.0  & 1RXS J184739.3$+$015732 &   1.30$\pm$ 0.10 &  30.0\\
1RXS J071506.6$-$702540 &   1.83$\pm$ 0.21 &  43.6 & 1RXS J184756.1$-$221938 &   2.79$\pm$ 0.59 &  31.6 \\
1RXS J072047.8$-$314705 &   0.74$\pm$ 0.12 &  52.4  & 1RXS J191824.6$+$595953 &   4.59$\pm$ 1.00 &  33.0 \\
1RXS J072320.0$-$274720 &   1.28$\pm$ 0.08 &  37.2  & 1RXS J192558.3$-$563344 &   7.71$\pm$ 0.69 &  50.0 \\
1RXS J072905.9$-$384839 &   5.27$\pm$ 0.85 &  40.0  & 1RXS J194343.1$+$500440 &   6.57$\pm$ 1.45 &  34.4\\
1RXS J080933.1$-$725909 &   1.23$\pm$ 0.28 &  30.8 & 1RXS J200905.6$-$602537 &   3.84$\pm$ 0.11 &  41.9 \\
1RXS J082335.2$-$252522 &   3.02$\pm$ 0.66 &  43.2  & 1RXS J201310.0$+$400222 &   6.67$\pm$ 0.74 &  53.6 \\
1RXS J082704.9$+$284411 &   4.81$\pm$ 0.80 &  40.0  & 1RXS J202400.4$-$422429 &   1.89$\pm$ 0.36 &  29.1 \\
1RXS J083151.0$-$534029 &   2.02$\pm$ 0.46 &  30.5  & 1RXS J211244.1$+$500616 &   1.20$\pm$ 0.06 &  39.8 \\
1RXS J084104.2$+$032118 &   3.55$\pm$ 0.31 &  38.2 & 1RXS J211652.9$+$735037 &   8.56$\pm$ 0.71 &  54.7\\
1RXS J084548.9$+$485241 &   1.47$\pm$ 0.15 &  39.5 & 1RXS J212458.2$+$282553 &   1.63$\pm$ 0.62 &  53.0\\
1RXS J091422.0$+$021916 &   1.01$\pm$ 0.23 &  30.0 & 1RXS J212626.8$+$192224 &   1.68$\pm$ 0.13 &  35.0\\
1RXS J091657.6$-$194613 &  11.03$\pm$ 1.73 &  56.4 & 1RXS J212743.5$-$221145 &   3.73$\pm$ 0.84 &  49.8  \\
1RXS J094021.7$+$502116 &   1.77$\pm$ 0.52 &  36.2 &  1RXS J215453.3$-$302935 &   0.78$\pm$ 0.24 &  29.4\\
1RXS J095752.3$+$852935 &   6.32$\pm$ 0.74 &  52.0 &  1RXS J215621.5$-$543820 &   2.72$\pm$ 0.16 &  45.8 \\
1RXS J101628.3$-$052026 &   4.85$\pm$ 0.41 &  54.8 & 1RXS J215635.1$-$414220 &   4.36$\pm$ 0.51 &  48.2 \\
1RXS J102405.9$+$262113 &   3.27$\pm$ 0.85 &  37.0 & 1RXS J221030.3$-$300546 &   1.01$\pm$ 0.24 &  28.8  \\
1RXS J102444.5$-$302102 &   1.83$\pm$ 0.36 &  34.8 &  1RXS J224952.6$+$583430 &  16.40$\pm$ 1.61 &  59.4 \\
1RXS J102945.3$+$450717 &   1.34$\pm$ 0.35 &  35.4 & 1RXS J231221.6$+$104657 &   1.28$\pm$ 0.09 &  60.1  \\
1RXS J103210.2$+$532941 &   2.18$\pm$ 0.06 &  47.0 & 1RXS J232431.0$-$544148 &   2.21$\pm$ 0.19 &  44.8  \\
\enddata
\end{deluxetable*}

\end{document}